\documentclass{aip-cp}

\usepackage[numbers]{natbib}
\usepackage{rotating}
\usepackage{graphicx}
\usepackage{enumitem}

\begin{document}

\title{Fundamental Physics With\\Cosmic High-Energy Gamma Rays}

\author[aff1,aff2]{Alessandro De Angelis}

\affil[aff1]{INFN Padova, via Marzolo 8, I-35141 Padova, and INAF, Italy.}
\affil[aff2]{Universit\`a di Udine, Via delle Scienze 208, Udine, Italy; LIP/IST, Lisboa, Portugal.}

\maketitle

\begin{abstract}
High-energy photons (above the MeV) are a powerful probe for astrophysics and for fundamental physics
under extreme conditions. During the recent years, our knowledge of the high-energy gamma-ray sky has impressively progressed thanks to the advent of new detectors
for cosmic gamma rays, at ground (H.E.S.S., MAGIC, VERITAS, HAWC) and in space (AGILE, Fermi).  This presentation reviews the present status of 
the studies of fundamental physics problems with high-energy gamma rays, and discusses  the expected experimental
developments.
\end{abstract}

\section{INTRODUCTION}

High-energy (HE) photons, above the MeV, are characteristic of hard process well above the thermal equilibrium. The study of this energy band unveils the hart of the extreme Universe, and allows accessing to some phenomena inaccessible to accelerators.

Among the broad spectrum of scientific opportunities offered by the observation of the sky above the MeV energies,
 some  may have a relevant
impact in fundamental physics and in cosmology.
 
 This presentation summarizes some of the studies in fundamental physics with HE photons.
 
 The typical instruments used to detect HE photons are:
 
\begin{itemize}
\item in the energy region from the MeV to some 30 GeV, instruments in space -- the atmosphere makes it impossible to observe photons of this energy at ground. The pair-production telescope like AGILE and Fermi start being effective at energies above some 30 MeV; the energy region below 30 MeV requires the detection of the Compton scattering, and present detectors in the MeV range have very limited sensitivity and tracking capability.
\item in the energy region above $~30$ GeV (the so-called Very-High-Energy, VHE, region), observation at ground becomes possible. The Imaging Air Cherenkov Telescopes (IACT) H.E.S.S., MAGIC and VERITAS  \cite{ourbook} allow a good coverage of the energy region up to a few TeV, albeit with a limited field-of-view.
Above a few hundred GeV, direct sampling of secondary particles in the atmospheric showers becomes possible. The large Cherenkov Telescope Array CTA is in construction.  Extensive air shower (EAS) detectors, such as  
HAWC  presently in operation, 
are large arrays of detectors sensitive to charged secondary particles generated in the  showers.
They have a high duty cycle and a large field-of-view, but a relatively low sensitivity.
\end{itemize}

\begin{figure}[ht]
\centering
\includegraphics[width=0.7\textwidth]{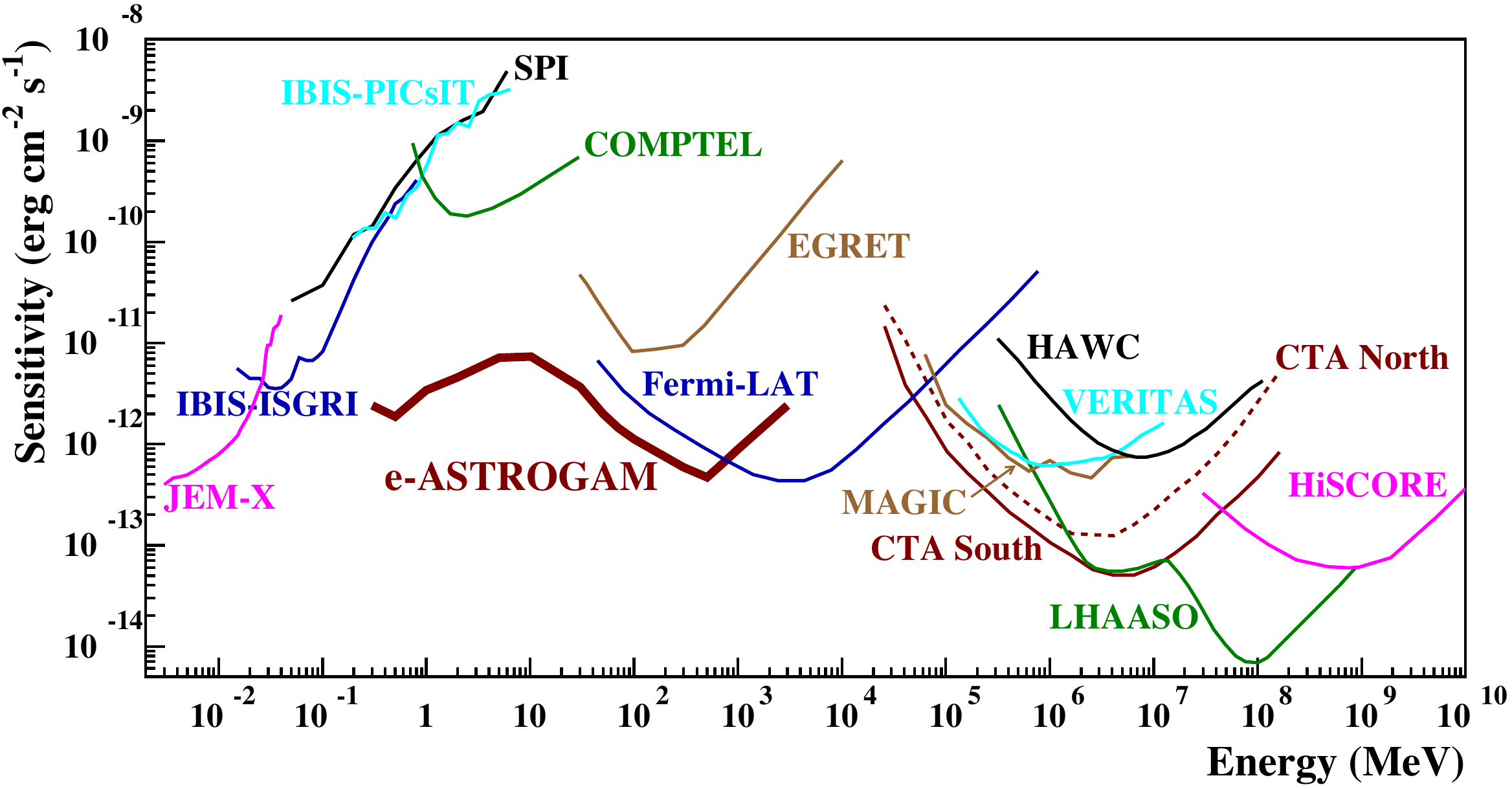}
\caption{{Point source continuum sensitivity of different X- and $\gamma$-ray instruments. The curves for INTEGRAL/JEM-X, IBIS (ISGRI and PICsIT), and SPI are for an observing time $T_{\rm obs}$ = 1 Ms. The COMPTEL and EGRET sensitivities are given for the  time accumulated during the  duration of the CGRO mission  ($T_{\rm obs} \sim $ 9 years). The Fermi/LAT sensitivity is for a high Galactic latitude source over 10 years. For MAGIC, VERITAS, and CTA, the sensitivities are given for $T_{\rm obs}$ = 50 hours. For HAWC $T_{\rm obs}$ = 5 yr, for LHAASO $T_{\rm obs}$ = 1 yr, and for HiSCORE $T_{\rm obs}$ = 1000 h. The e-ASTROGAM sensitivity is for an effective exposure of 1 year for a source at high Galactic latitude.}
\label{fig:Sensitivities}}
\end{figure}

\begin{figure} [\sidecaptionrelwidth]
\centering
\includegraphics[width=0.7\textwidth]{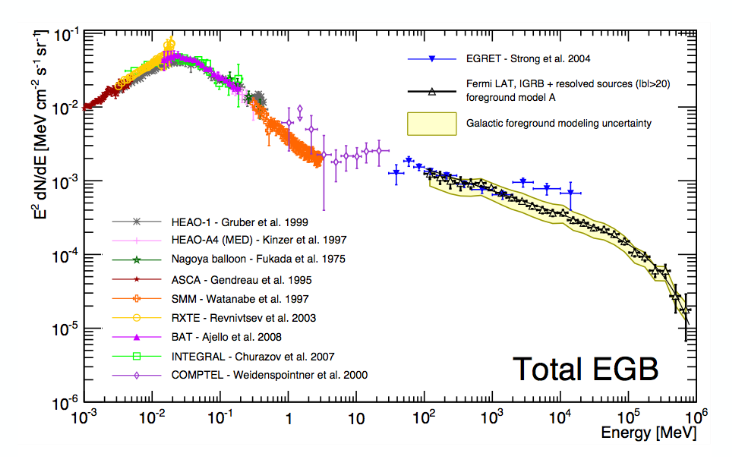}
\caption{Compilation of the measurements of the total extragalactic gamma-ray intensity between 1 keV and 820 GeV \cite{FermiEGB}, {with different components from current models}; the contribution from MeV blazars is largely unknown.
  The semi-transparent band indicates the energy region in which e-ASTROGAM will dramatically improve on present knowledge.\label{fig:egb}}
\end{figure}

A compilation of the sensitivities of different present and future gamma-ray detectors is shown in Figure \ref{fig:Sensitivities}.
A compilation of the results on the extragalactic population of HE photons is shown in Figure \ref{fig:egb}.

Let us now review some of the subjects related to fundamental physics for which high-energy comic photons can be important.

\section{INDIRECT SEARCH FOR DARK MATTER}

Evidence for departure of cosmological motions from the predictions of Newtonian dynamics 
based on visible matter, are well established 
-- from galaxy scales to galaxy-cluster scales to cosmological scales. 
Such departures could be interpreted as due to the presence of an undectected matter, called Dark Matter (DM).  
A remarkable
agreement of a diverse set of astrophysical data indicates that the energy budget of DM could be ~26\% of the 
total energy of the Universe, compared to some 4\% due  to ordinary matter.

DM particle candidates should be weakly interacting with ordinary matter (and hence neutral). 
The  theoretically favored ones  are heavier than the proton, and they are dubbed weakly interacting 
massive particles (WIMPs). WIMPs should be long-lived enough to have survived from their 
decoupling from radiation in the early universe into the present epoch; they should interact among them, annihilating in pairs. 
 WIMP candidates have been proposed within theoretical 
frameworks mainly motivated by extensions of the Standard Model (SM) of particle physics (e.g., the 
R-parity conserving supersymmetry, SUSY). Among current WIMP candidates, 
the neutralino $\chi,$ which is the lightest SUSY particle, is the most popular candidate. Its relic 
density is compatible with cosmological bounds, if its mass is of the order of 10-100 GeV. In what follows we shall call $\chi$ a generic WIMP.

For a WIMP mass O(100 GeV), if WIMPS were in thermal equilibrium and decoupling,
\begin{equation}\label{eq:relicdensity}
{<|v_\chi| \sigma_A>} \sim 3 \times 10^{-26} {\rm{cm^3/s}}  
\end{equation}
where $v_\chi$ is the
relative velocity between the two WIMPs and $\sigma_A$ is the annihilation cross section. The quantity $|v|\sigma$ is expected not too change in an appreciable way during the evolution of the Universe. Thus, the value ${<|v_\chi| \sigma_A>} = 3 \times 10^{-26} {\rm{cm^3/s}}$ is a benchmark in the search for dark matter particles.

The $\gamma$-ray flux from the annihilation of dark matter particles of mass $m_{\chi}$ can be 
expressed as the product of a particle physics component times an astrophysical component (this last being often called the {\em{J-factor}}):
\begin{equation}
\frac{dN}{dE}\,=\frac{1}{4\pi}\,\underbrace{\frac{\langle
\sigma
v\rangle}{m^2_{DM}}\,\frac{dN_{\gamma}}{dE}}_{Particle\,
Physics}\,\times\,\underbrace{\int_{\Delta\Omega-l.o.s.} dl(\Omega) \, \rho^2_{\chi}}_{Astrophysics} \ . \label{eq:dm}
\end{equation}
The particle physics factor contains $\langle \sigma v\rangle$, the velocity-weighted 
annihilation cross section (there is indeed a possible component from cosmology in $v$), and 
$dN_{\gamma}/dE$, the differential $\gamma$-ray spectrum summed over the final states with 
their corresponding branching ratios.
The astrophysical part corresponds to the squared density of the dark matter distribution 
integrated over the line of sight ({\it l.o.s.}) in the observed solid angle.

A similar expression to Equation \ref{eq:dm} holds for the production of antimatter in $\chi\chi$ annihilations.

It is clear that the expected flux of photons (and of antimatter) from dark matter annihilations, and thus its 
detectability, depends crucially on the knowledge of the annihilation cross section $\sigma$ (which even within SUSY has 
uncertainties of one-two orders of magnitude for a given WIMP mass) and of $\rho_{\chi}$, which 
is even more uncertain, and enters squared in the calculation.

\subsection{Photon signatures}

WIMP self-annihilation can generate 
$\gamma$-rays through several processes. Most distinctive are those that result in mono-energetic 
spectral lines, $\chi\chi$$\rightarrow$$\gamma\gamma$, $\chi\chi$$\rightarrow$$\gamma$$Z$. However, in most models the processes only take place through 
loop diagrams; hence the cross sections for such final states are quite suppressed, and the lines 
are weak and experimentally challenging to observe. A continuum $\gamma$-ray spectrum can also be 
produced through the fragmentation and cascades of most other annihilation products. The resulting 
spectral shape depends on the dominant annihilation modes, whereas 
the normalization depends on the WIMP's velocity-averaged annihilation cross section as well 
as on the DM density profile. 

\begin{figure}
\includegraphics[width=0.4\linewidth]{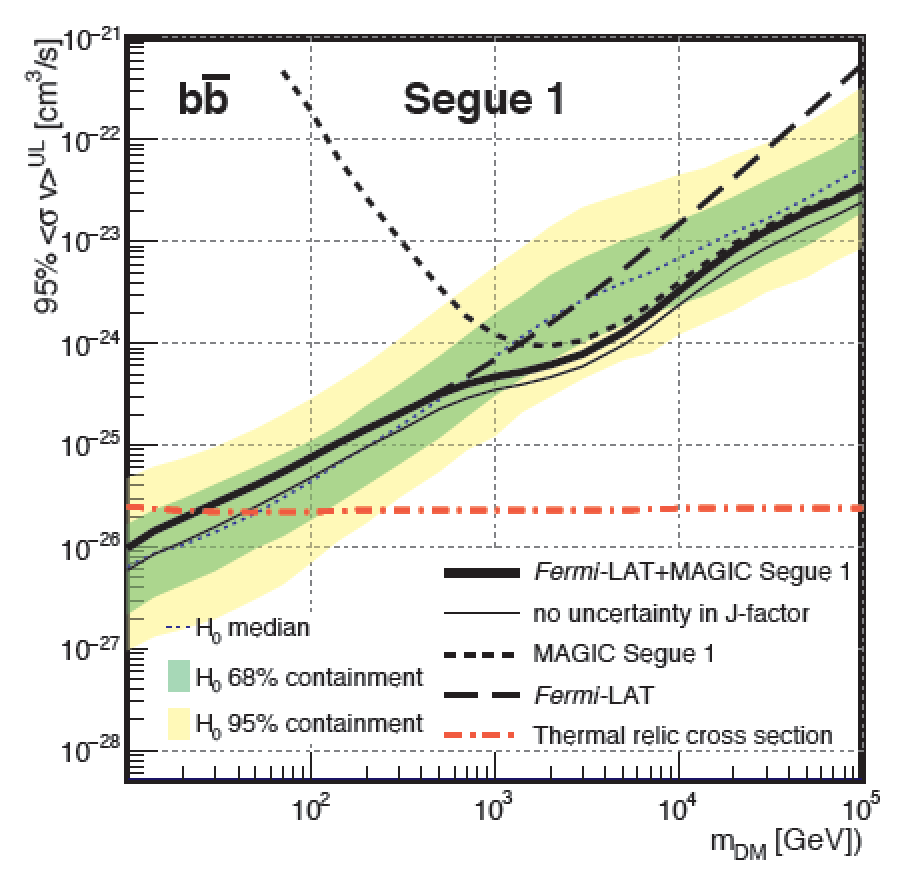}
\vspace*{-2mm}
\includegraphics[width=0.4\linewidth]{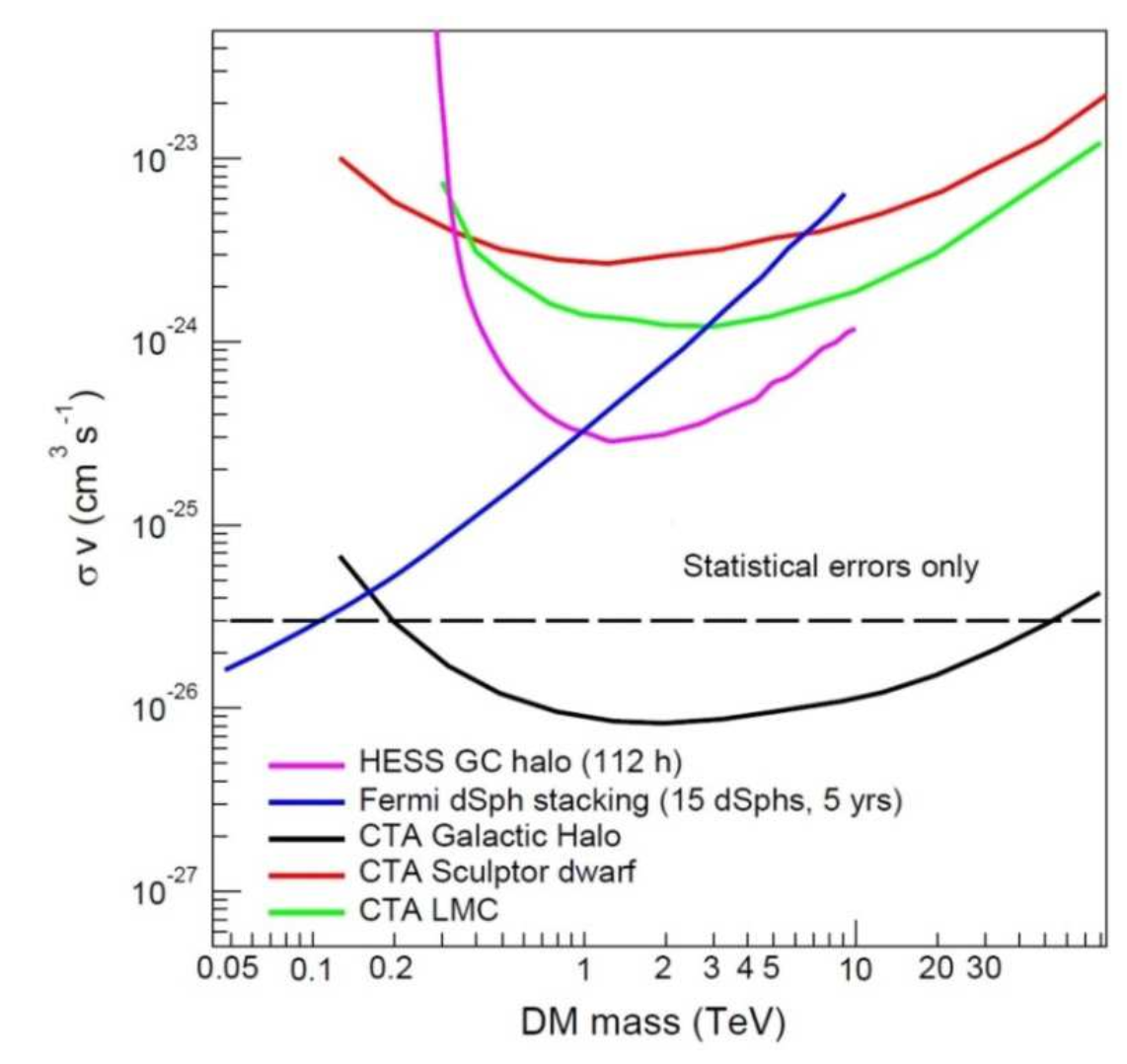}
\caption{Left panel: 95\% CL upper limits on the thermally-averaged cross-section for DM particles annihilating into
$b\bar{b}$
 pairs. Thick solid lines
show the limits obtained by combining Fermi-LAT observations  with MAGIC observations of
Segue 1. Dashed lines show the observed individual MAGIC (short dashes) and Fermi-LAT (long dashes)
limits. Right panel: compilation of present results from gamma-ray telescopes. Right: Potential of CTA. \label{compidm}}
\end{figure}

The main advantage of searches using photons is that one can point the expected sources. Which targets are the best \cite{carr} to search for DM with gamma rays?
\begin{itemize}
\item The center of the Milky Way has most likely the highest DM concentration in the Galaxy.
Because of the
 rich field of  high energy 
gamma-ray astrophysical sources in the region, the 
searches is often focused on the Galactic halo 
in the Inner Galaxy: even excluding the very 
central  region,  the  total  mass  of  DM  in  the  galactic  halo  together  with  its 
proximity to Earth make it the most promising source for DM searches with IACTs. The main
inconvenience  of  this  target,  however,  is  the  fact 
that  there  are 
astrophysical  backgrounds  from  various  sources  which  must  be  understood,  even  with 
the very central region excluded from the analysis.
Studying these backgrounds will require a multi wavelength high-energy analysis, possibly starting from the MeV region.
\item The dwarf spheroidal galaxies (dSphs) of the Local 
Group could give a clear and 
unambiguous detection of dark matter \cite{17}. They are
 gravitationally bound objects and 
are  believed  to  contain  up  to  three orders of magnitude  more  mass  in  dark  matter  than  in  visible 
matter, making them widely discussed as potential targets. Since you do not expect gamma-ray emission apart from DM self-annihilation, observations are background free. Their  J-factors integrated over the small 
source  have  less  dependence  on  the  DM  profile  assumed  than  the  extended  sources. 
Neither  astrophysical  gamma-ray  sources  (supernova 
remnants,  pulsar  wind  nebulae...) 
nor gas acting as target material for cosmic rays, 
have been observed in these systems.  
Due  to  the  larger  available  sample  of  spectroscopic
ally  measured  stars,  the 
``classical''  dwarf  galaxies  such  as  Draco,  Ursa  Minor,
  Carina,  and  Fornax  have  the 
significantly  smaller  uncertainties  on  the  J-factor
  than  the  ultra-faint  dwarf  galaxies.  However  several  of  the  ultra-faint  galaxies  (e.g.
  Segue  1,  Ursa  Major  II,  and                                               
Reticulum  II)  have  J-factors  which  are  larger  than 
the  J-factors  of  the  best  classical 
dwarfs. 
Examples  of  the  sensitivity  which  could  be  obtained
  by  observations  of  a  classical 
dwarf galaxy are  shown in Figure \ref{compidm}.
\end{itemize}
In addition to the above targets, structure formation predicts gravitationally bound dark matter clumps down to
much lower masses than observed for dSph galaxies between $10^{-12} M_\odot$ and $10^{-3} M_\odot$  for typical WIMP
scenarios \cite{9}. 

A compilation of present data and of the expected CTA results is shown in Figure \ref{compidm}. The observational strategy proposed for the CTA Dark Matter programme is
focused first on collecting a significant amount of data on the centre of the Galactic
Halo. Complementary observations of a dSph galaxy will be conducted to extend the
dark matter searches. The Galactic Halo and Large Magellanic Cloud are valuable
targets both for dark matter searches and studies of non-thermal processes in
astrophysical sources. Data will be searched for continuum emission and line features,
and strategies will be adopted according to findings.

\subsection{Antimatter signatures}

%
%
%
PAMELA and AMS-02 have observed
an anomalous positron abundance in cosmic
radiation (see \cite{creview} for a review). 
Among possible explanations, DM annihilation
is especially interesting.

Separation of electrons from positrons in Cherenkov telescopes can be achieved by IACTs.
The Moon produces a 0.5$^\circ$-diameter hole in the isotropic CR flux, which is shifted by the Earth magnetosphere depending on the momentum and charge of the particles. Below a few TeV, the positron and electron shadows are shifted by more than $0.5^\circ$ each side of the Moon due to the geomagnetic field.

Among the possible explanations of the excess is also the production in nearby, presently unknown, astrophysical sources (pulsars for example), Gamma-ray detectors measuring the such sources could thus be the key to understand if the positron excess is due to new physics. Fermi has given hints for possible explanations of the excess as due to pair production in nearby pulsars; experiments in the MeV-GeV range are needed to clarify \cite{deaastro,astrogam,compair}.

\section{PHOTON PROPAGATION, PHOTON BACKGROUND AND FUNDAMENTAL PHYSICS}

Electron-positron $(e^-e^+)$ pair production in the interaction of VHE photons off extragalactic background photons is a source of opacity of the Universe to $\gamma$-rays whenever the corresponding photon mean free path is of the order of the source distance, or smaller.


The dominant process for the absorption is pair-creation 
\[ \gamma + \gamma_{\,\texttt{\scriptsize background}} \rightarrow{} e^+ + e^- \, ; \]
the process is kinematically allowed for
\begin{equation} 
\label{eq.sez.urto01012011}
\epsilon > {\epsilon}_{\rm thr}(E,\varphi) \equiv \frac{2 \, m_e^2 \, c^4}{ E \left(1-\cos \varphi \right)}~,
\end{equation}
where $\varphi$ denotes the scattering angle, $m_e$ is the electron mass, $E$ is the energy of the incident photon and $\epsilon$
is the energy of the target (background) photon. Note that $E$ and $\epsilon$ change along the line of sight in proportion of $(1 + z)$ because of the cosmic expansion. The corresponding  cross-section, computed by Breit and Wheeler in 1934, is
\begin{equation} 
\sigma_{\gamma \gamma}(E,\epsilon,\varphi)  = \frac{2\pi\alpha^2}{3m_e^2}  W(\beta) 
                                                         \simeq 1.25 \cdot 10^{-25} \,  W(\beta) \,  {\rm cm}^2~, \label{eq:bhe} 
\end{equation}
with
\[ W(\beta) = \left(1-\beta^2 \right) \left[2 \beta \left( \beta^2 -2 \right) 
+ \left( 3 - \beta^4 \right) \, {\rm ln} \left( \frac{1+\beta}{1-\beta} \right) \right] \, . \]
The cross-section depends on $E$, $\epsilon$ and $\varphi$ only through the  speed $\beta$ -- in natural units -- of the electron and of the positron in the center-of-mass
\begin{equation} 
\label{eq.sez.urto01012011q}
\beta(E,\epsilon,\varphi) \equiv \left[ 1 - \frac{2 \, m_e^2 \, c^4}{E \epsilon \left(1-\cos \varphi \right)} \right]^{1/2}~, 
\end{equation}
and Eq. (\ref{eq.sez.urto01012011}) implies that the process is kinematically allowed for ${\beta}^2 > 0$. The cross-section $\sigma_{\gamma \gamma}(E,\epsilon,\varphi)$ reaches its maximum ${\sigma}_{\gamma \gamma}^{\rm max} \simeq 1.70 \cdot 10^{- 25} \, {\rm cm}^2$ for $\beta \simeq 0.70$. Assuming head-on collisions  ($\varphi = \pi$), it follows that $\sigma_{\gamma \gamma}(E,\epsilon,\pi)$ gets maximized for the background photon energy 
\begin{equation} 
\label{eq.sez.urto-0}
\epsilon (E) \simeq \left(\frac{500 \, {\rm GeV}}{E} \right) \, {\rm eV}~,
\end{equation}
where $E$ and $\epsilon$ correspond to the same redshift. 
 For an isotropic background of photons, the cross-section is maximized for background photons of energy:
 \begin{equation} 
 \label{eq.sez.urto-1}
\epsilon (E) \simeq \left(\frac{900 \, {\rm GeV}}{E} \right) \, {\rm eV}~.
\end{equation}
 
 Explicitly, the situation can be summarized as follows.
\begin{itemize}
\item For $10 \, {\rm GeV} \leq E < 10^5 \, {\rm GeV}$ the EBL plays the leading role in the absorption. In particular, 
for $E \sim 10 \,{\rm GeV}$ $\sigma_{\gamma \gamma}(E,\epsilon)$ -- integrated over an isotropic distribution of background photons -- is maximal for $\epsilon \sim 90 \, {\rm eV}$, corresponding to far-ultraviolet soft photons, whereas for $E \sim 10^5 \,{\rm GeV}$ $\sigma_{\gamma \gamma}(E,\epsilon)$ is maximal for $\epsilon \sim 9  \, {\rm meV}$, corresponding to soft photons in the far-infrared.
\item For $10^5 \, {\rm GeV} \leq E < 10^{10} \, {\rm GeV}$ the interaction with the CMB becomes dominant.
\item For $E \geq 10^{10} \, {\rm GeV}$ the main source of opacity of the Universe, is the radio background.
\end{itemize}

From the cross section in Equation (\ref{eq:bhe}), neglecting the expansion of the Universe one can compute a mean free path; for energies smaller than some 10 GeV this is larger than the Hubble radius, but it becomes comparable with the distance of observed sources at energies above 100 GeV.
%

The attenuation suffered by observed VHE spectra can thus be used to derive constraints on the 
EBL density.
Specifically, the probability $P$ for a photon of observed energy $E$ to survive absorption along 
its path from its source at redshift $z$ to the observer plays the role of an attenuation 
factor for the radiation flux, and it is usually expressed in the form:
\begin{equation} \label{eq:flux.tau}
P = e^{-\tau(E,z)} \, .
\end{equation}
The coefficient $\tau(E,z)$ is called {\em optical depth}. \index{optical depth} \index{high-energy gamma-rays!optical depth} 

To compute the optical depth of a photon as a function of its observed energy~$E$ and the 
redshift $z$ of its emission one has to take into account the fact that the energy $E$ of a 
photon scales with the redshift~$z$ as $(1+z)$; thus when using 
Equation ~\ref{eq:bhe} we must treat the energies as function of~$z$ and evolve 
$\sigma\big(E(z),\epsilon(z),\theta\big)$ for 
$E(z)= (1+z)E$ and $\epsilon(z)=(1+z)\epsilon$,
where $E$ and $\epsilon$ are the energies at redshift $z=0$.
The optical depth is then computed by convolving the photon number density 
of the background photon field with the cross section between the incident $\gamma$-ray and 
the background target photons, and integrating the result over the distance, the scattering 
angle and the energy of the (redshifted) background photon:
\begin{equation}  
\tau(E,z) = 
\int_{0}^{z} dl(z)\
\int_{-1}^{1}\ d\cos \theta \frac{1-\cos \theta}{2} 
 \int_{\frac{2(m_e c^2)^2}{E(1-\cos\theta)}}^{\infty} d\epsilon(z)\
n_{\epsilon}\big(\epsilon(z),z\big) \ \sigma(E(z),\epsilon(z),\theta) \label{eq:comptau}
\end{equation}
where
$\theta$ is the scattering angle,
$n_{\epsilon}\big(\epsilon(z),z\big)$ is the density for photons of energy $\epsilon(z)$ at 
the redshift $z$, and $l(z) = c\ dt(z)$ is the distance as a function of the redshift, defined 
by
\begin{equation}
\label{eq:padmanabhan-diff}
\frac{dl}{dz} \ = \ \frac{c}{H_0} 
\frac{1}{(1+z) \left[ (1+z)^2 (\Omega_M\,z+1) - \Omega_{\Lambda}\,z(z+2) \right]^{\frac{1}{2}} } \, .
\end{equation}
In the last formula  $H_0$ is the Hubble constant, $\Omega_M$ is the matter density (in units 
of the critical density, $\rho_{\rm c}$) and $\Omega_{\Lambda}$ is the ``dark energy'' density (in 
units of $\rho_{\rm c}$); therefore, since the 
optical depth depends also on the cosmological parameters, its determination constrains the 
values of the cosmological parameters if the cosmological emission of galaxies 
is known.


The energy dependence of $\tau$ leads to appreciable modifications of the observed source 
spectrum (with respect to the spectrum at emission) even for small differences in~$\tau$, due to the 
exponential dependence described in Equation ~(\ref{eq:flux.tau}).
Since the optical depth (and consequently the absoption coefficient) increases with energy, 
the observed flux results steeper than the emitted one.

\begin{figure}
\centering
\includegraphics[width=.36\textwidth]{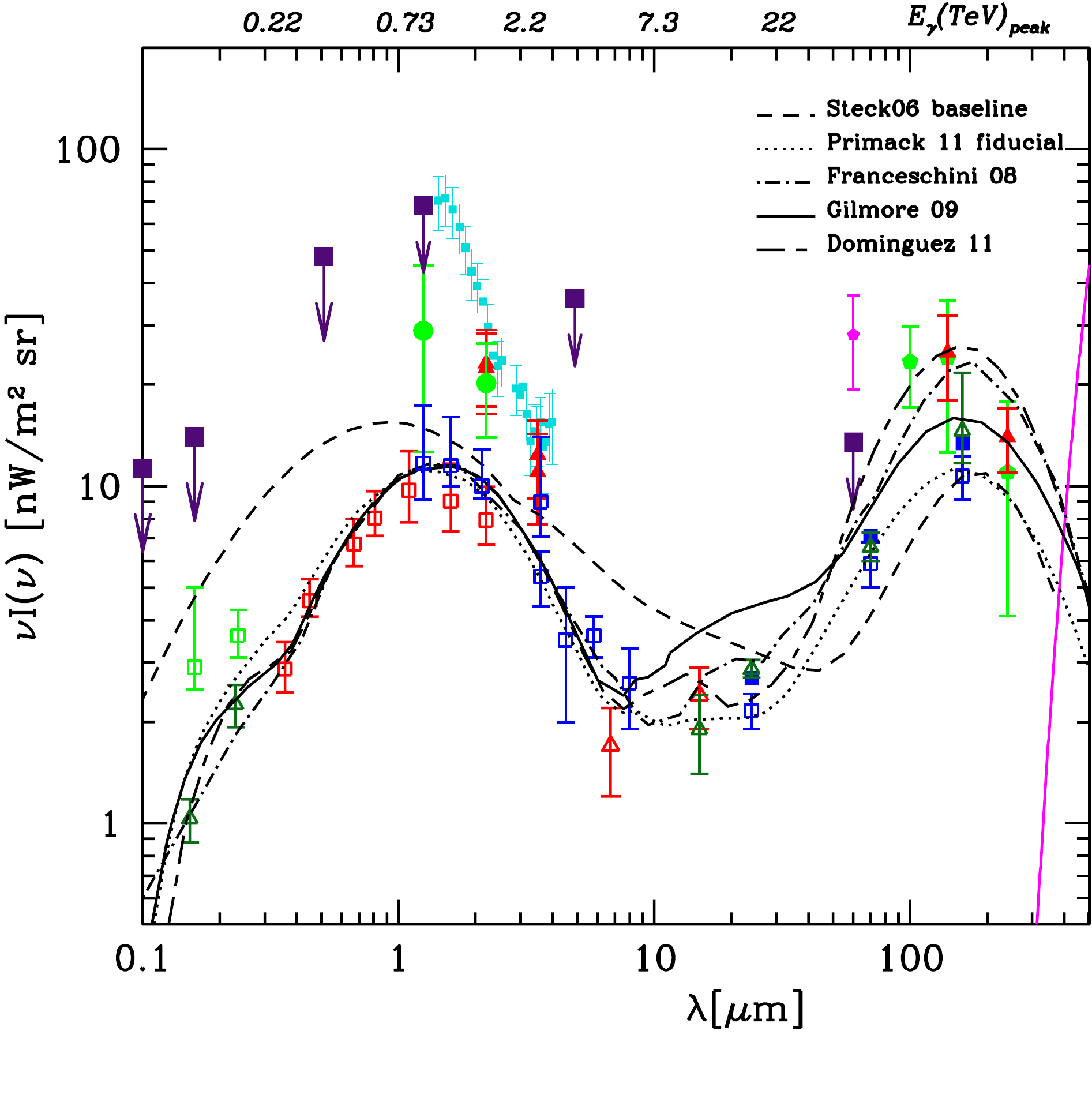}
 \includegraphics[width=.47\textwidth]{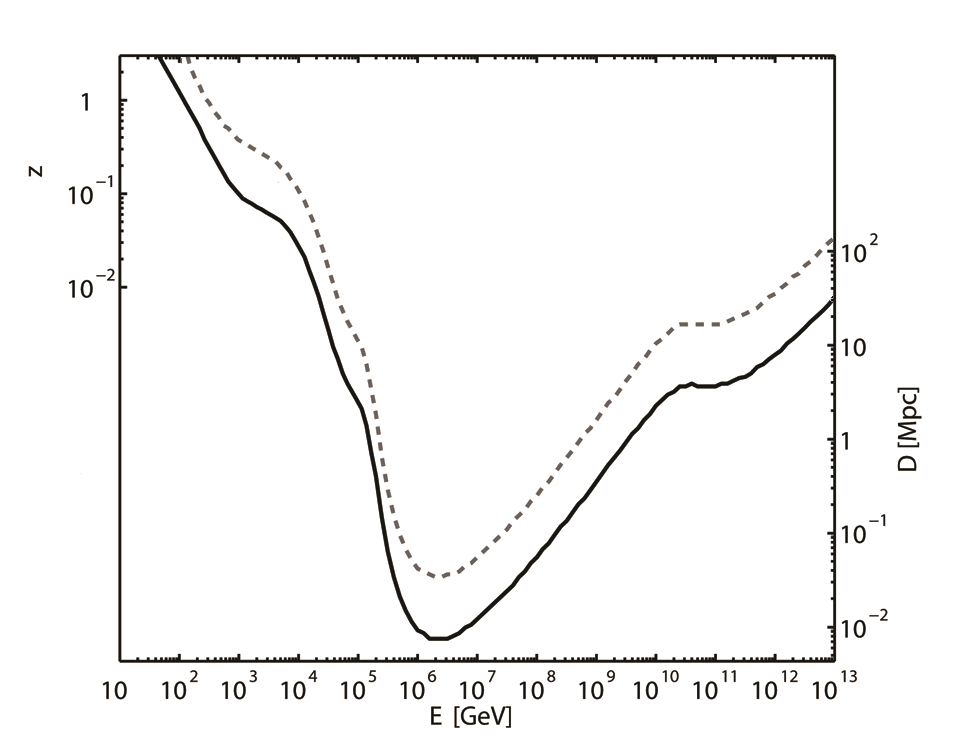}
\caption{\label{fig:CoppiAharonian}
Left:  Spectral energy distribution of the EBL as a function of the wavelength.  Open symbols correspond to lower limits from galaxy counts
while filled symbols correspond to direct estimates.  The curves show a sample of different recent EBL models, as labeled. 
On the upper axis  the TeV energy corresponding to the peak of the $\gamma\gamma$ cross section is plotted. From \cite{luigi}.
 Right:
\label{fig:gr-horizon}
Curves corresponding to the gamma-ray horizon $\tau(E,z)=1$ (lower) and to a survival probability of $e^{-\tau(E,z)}$ = 1\% (upper). Adapted from
\cite{noimras}.}
\end{figure}

The {\em horizon}  or {\em attenuation edge}  \index{gamma-ray!horizon} \index{high-energy gamma-rays!attenuation edge}  
 for a photon of energy $E$ is defined as the distance 
corresponding to the redshift~$z$ for which $\tau(E,z)=1$, that gives an attenuation by 
a factor $1/e$ (see Figure \ref{fig:gr-horizon}).

Other interactions than the one just described might change our picture of the attenuation 
of $\gamma$-rays, and they are presently subject of thorough studies, since the present data 
on the absorption of photons are hardly compatible with the pure QED picture: from the observed luminosity of VHE photon sources,
the Universe appears to be more transparent to $\gamma$ rays than expected.
One speculative explanation could be that $\gamma$-rays might  transform into sterile or quasi-sterile particles (like, for example, axions); this would increase the transparency by effectively decreasing the path length. A more detailed discussion will be given later in this Section.

%

\subsection{The transparency of the Universe: search for axion-like particles}

Some experimental indications exist, that the Universe might be more transparent to gamma-rays than computed on the basis of EBL as presently known. 

Several models have been proposed in the literature to estimate the spectral energy density (SED) of the soft background (EBL);
since they are based on suitable experimental evidence (e.g., deep galaxy counts), all models 
yield consistent results, so that the SED of the EBL is fixed to a 
very good extent. Basically, the latter reproduces the SED of star-forming galaxies, which is 
characterised by a visible/ultraviolet hump due to direct emission from stars and by an infrared 
hump due to the emission from the star-heated warm dust that typically hosts the sites of star 
formation.

\begin{figure}
\centering
\includegraphics[width=.6\textwidth]{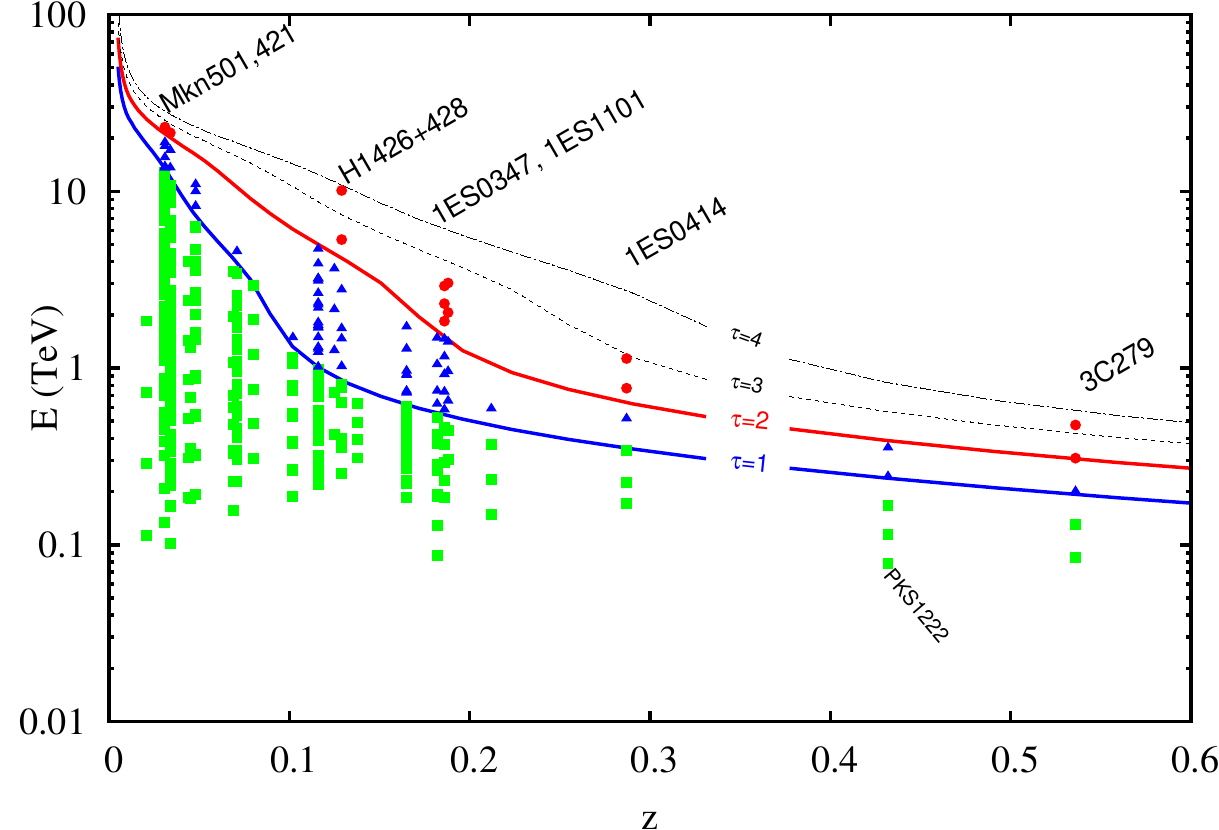}
\caption{\label{fig:horns}
For each individual spectral measurement, the corresponding value of $z$ and $E$ are marked in this diagram. The iso-contours for $\tau = 1,\, 2,\, 3,\, 4$ calculated using a minimum EBL model  are overlaid \cite{horns}.}
\end{figure}

\begin{figure}
\centering
\includegraphics[width=.6\textwidth, height=4cm]{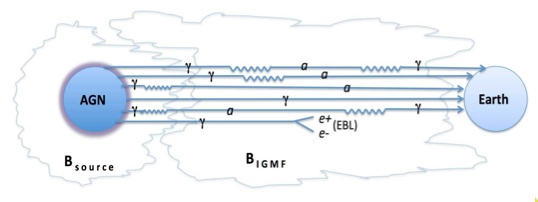}
\includegraphics[width=.4\textwidth, height=4cm]{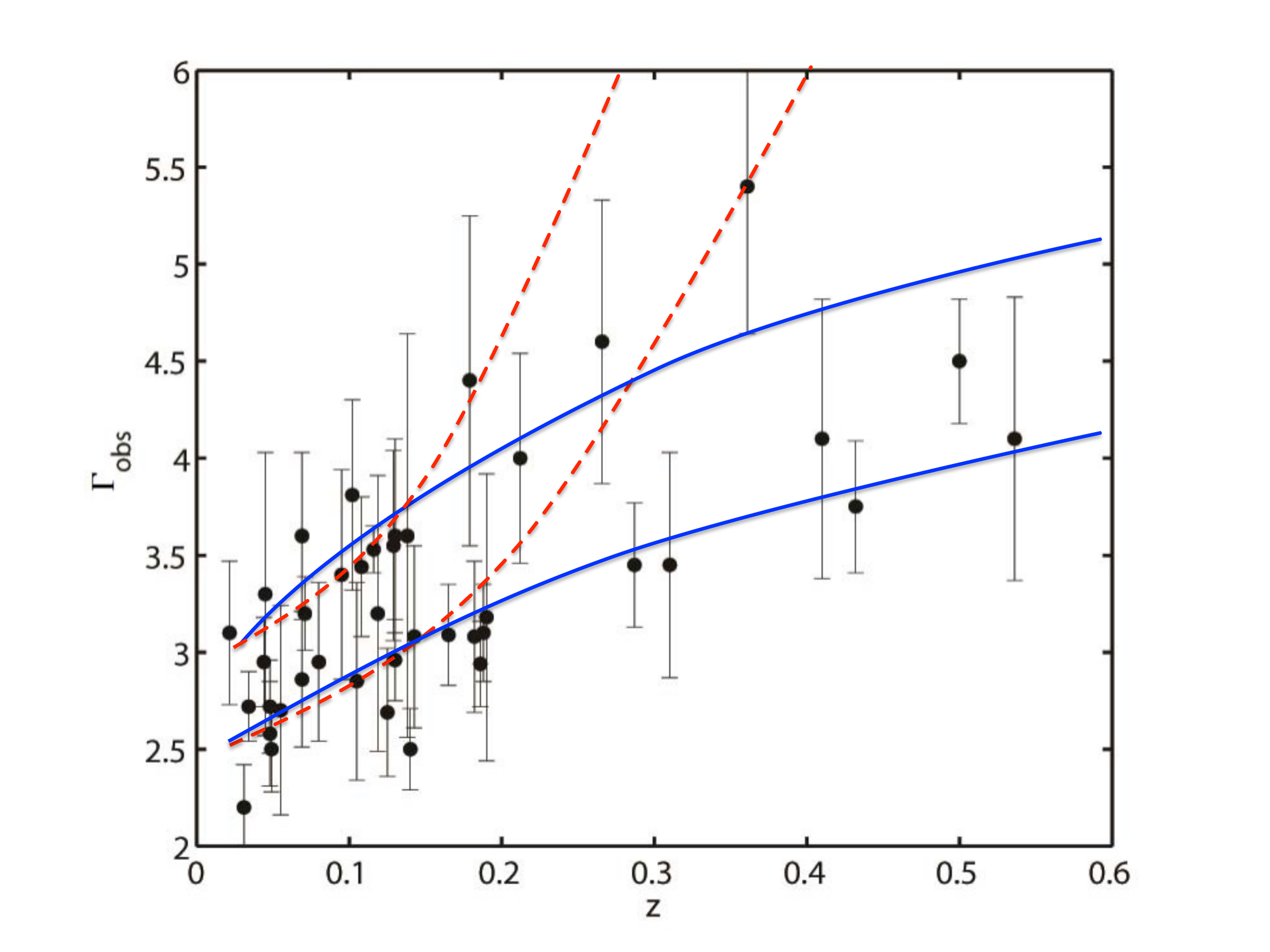}
\caption{\label{fig:darma}
Left: Illustration of gamma-ray propagation in the presence of oscillations between gamma-rays and axion-like particles. From \cite{sanchez}.
\label{fig:spectral}
Right: The observed values of the spectral index for all blazars detected   in VHE; superimposed   is the
predicted behavior of the observed spectral index from a source at constant intrinsic spectral index within two different scenarios. In the first
one (area between the two dotted lines) $\Gamma$ is computed from EBL absorption; in the second (area between the two solid lines) it is
evaluated including also the photon-ALP oscillation \cite{ourbook}.}
\end{figure}

However, the Universe looks more transparent than expected -- this is called the ``EBL-crisis''. Basically two experimental evidences support this conjecture:
\begin{itemize}
\item When for each SED of high-$z$ blazars, the data points observed in the optically thin regime $(\tau < 1)$ are used to fit the VHE spectrum in optically thick regions, points at large attenuation are observed (Figure \ref{fig:horns}). This violates the current EBL models, strongly based on observations, at some $5\sigma$.
\item The energy dependence of the gamma opacity $\tau$  leads to appreciable modifications of the observed
source spectrum with respect to the spectrum at emission, due to the exponential
decrease  of $\tau$ on energy in the VHE gamma region. One would expect naively that the spectral index of blazars at VHE would increase with distance: due to absorption, the SED of blazars should become steeper at increasing distance. This phenomenon has not been observed (Figure \ref{fig:spectral}).
\end{itemize}

Among the possible explanations, a photon mixing with axion-like particles (ALPs), \index{ALP (axion-like-particle)} \index{WISP} predicted by
several extensions of the Standard Model, can fix the EBL crisis, and obtain compatibility on the horizon calculation.  Since ALPs are characterised by a coupling to two photons, in the presence of an external magnetic field $B$  photon-ALP oscillations can show up. Photons are supposed to be emitted by a blazar in the usual way;  some of them can turn into ALPs, either 
in the emission region, or during their travel. Later, some of the produced ALPs can convert back into photons (for example in the Milky Way, which has a 
relatively large magnetic field) and ultimately be detected. In empty space this would obviously produce a flux dimming; remarkably enough, due to the EBL such a double conversion can make the observed flux considerably larger than in the standard situation:  in fact, ALPs do not undergo EBL absorption (Figure \ref{fig:darma}).

We concentrate now on the photon transition to ALP in the intergalactic medium. The 
probability of photon-ALP mixing depends on the value and on the structure of the cosmic magnetic fields, largely unknown.

Both the strength and the correlation length of the cosmic magnetic fields do influence the 
calculation of the $\gamma \to a$ conversion probability. In the limit of low conversion 
probability, if $s$ is the size of the typical region, the average probability $P_{\gamma 
\rightarrow a}$ of conversion in a region is
\begin{equation}
P_{\gamma \rightarrow a} \simeq 2 \times 10^{-3} \left( \frac{B_T}{\rm{1\, nG}}  \frac{\lambda_B}{\rm{1\, Mpc}}  
 \frac{g_{a\gamma\gamma}}{\rm{10^{-10}\, GeV^{-1}}}   \right) ^2 \, ,
\end{equation}
where $B_T$ is the transverse component of the magnetic field.

For a magnetic field of 0.1 nG -1 nG, and a cellular size structure $\lambda_B \sim 1 {\rm{Mpc}}-10$~Mpc, any ALP mass below 10$^{-10}$~eV, with a coupling 
such that $10^{11}$ GeV$ < M < 10^{13}$ GeV (well within the region experimentally allowed for mass and coupling) can explain the experimental results (Figure  \ref{fig:spectral}).

Another possible explanation for the hard spectra of distant blazars, needing a more fine tuning, is that line-of-sight interactions of cosmic rays with cosmic microwave background radiation  and EBL generate secondary gamma-rays relatively close to the observer.

\subsection{Lorentz Invariance Violation}

Mechanisms in which the absorption is changed through violation of the Lorentz 
invariance  are also under scrutiny; such models are particularly 
appealing within scenarios inspired by quantum gravity.

The variability of the AGN in the VHE region can provide information about possible violations 
of the Lorentz invariance in the form of a dispersion relation for light expected, for example, in some quantum gravity (QG)
models.

Lorentz invariance violation (LIV) \index{Lorentz!invariance violation}  at the $n-th$ order in energy can be heuristically incorporated in a perturbation to the relativistic Hamiltonian:
\begin{equation}\label{eq:spectral}
  E^2 \simeq  m^2c^4 + p^2c^2  \left[1-\xi_n\left( \frac{pc}{E_{\rm LIV,n}}\right)^n\right ] \, , 
\end{equation}
which implies that the speed of light ($m=0$)
could have an energy dependence. From the expression $v=\partial E/\partial p$,
 the modified dispersion relation of photons can be expressed by the leading term of the Taylor series as
 an energy-dependent light speed
\begin{equation}
 v(E) = \frac{\partial E}{\partial p}  \simeq c\left[1-\xi_n\frac{n+1}{2} \left( \frac{E}{E_{\rm LIV,n}}\right)^n \right], \label{eq:cspeed}
\end{equation}
where n=1~or n=2 corresponds to linear or quadratic energy dependence, and $s_n=\pm1$ is the sign of the LIV correction.
If $s_n=+1~(s_n=-1)$,  high-energy photons travel in vacuum slower~(faster) than  low energy photons. 

The   scale $E_{\rm LIV}$ at which the physics of space-time is expected to break down,  requiring modifications or the creation of a new paradigm to avoid singularity problems, is referred to as the ``QG energy scale'', and is  expected to be of the order of the 
Planck scale -- an energy $E_P = M_P c^2 \simeq 1.2 \times 10^{19}$ GeV -- or maybe lower, if new particles are discovered at an intermediate scale.


Because of the spectral dispersion, two GRB photons emitted simultaneously by the source would
arrive on Earth with a time delay~($\Delta t$) if they have different energies.
With the magnification of the cosmological distances of the GRBs and the high energies of these photons, the time delay ($\Delta t$) caused by the effect of Lorentz invariance violation could be measurable.
Taking account of the cosmological expansion and using Equation \ref{eq:cspeed}, we write the formula of the time delay as:
\begin{equation}\label{eq:Deltat}
  \Delta t=t_{h}-t_{l}=\xi_n\frac{1+n}{2H_0}
  \frac{E_{h}^n-E_{l}^n}{E_{LIV,n}^n}
  \int^z_0
  \frac{(1+z')^n dz'}{\sqrt{\Omega_m (1+z')^3+\Omega_\Lambda}}.
\end{equation}
Here, $t_{h}$ is the arrival time of the high-energy photon, and $t_{l}$ is the arrival time of the low energy photon,
with $E_{h}$ and $E_{l}$ being the photon energies measured at Earth. 

For small $z$, and at first order,
\[ t(E) \simeq d/c(E) \simeq \frac{z c_0}{H_0 c(E)} \simeq {z}T_H \left( 1 - \xi_1 \frac{E}{E_P} \right) \]
where $T_H = 1/H_0 \simeq 5 \times 10^{17}$ s is  Hubble's time.

%

AGN flares or GRBs can be used as experimental tools: they are fast, and photons arrive to us travel for long distances.

Mkn 501 (z = 0.034) had a spectacular flare  between May and July 2005; it could be analysed by the MAGIC telescope.  \index{MAGIC telescope}
The MAGIC data  showed a negative correlation between the 
arrival time of photons and their energy, yielding, if one 
assumes that the delay is due to linear quantum gravity effects, to an evaluation of $E_{\rm LIV} \sim 0.03~E_P$.
%
%
%
H.E.S.S.\ observations of the flare in PKS\,2155, however, evidenced no effect, allowing to set a \index{H.E.S.S. telescope}
lower limit $E_{\rm LIV} > 0.04~E_P$.

Lately, several GRBs observed by the Fermi satellite have been used to set more stringent limits. A problem, when setting limits, is that one does not 
know if photon emission at the source is ordered in energy; thus one has to make hypotheses -- for example, that QG effects can only increase the intrinsic dispersion.
The Fermi satellite derived strong upper limits at 95\% C.L. from the total degree of dispersion,   in the data of four  GRBs:
\[ E_{\rm LIV,1} > 7.6 E_P \, . \]

In most quantum gravity scenarios violations to the universality of the speed of light happen at order 
larger than 1:
$\Delta t \simeq \left( {E}/{E_{\rm LIV}} \right) ^\nu$ with $\nu > 1$.
In this case the VHE detectors are even more sensitive with respect to other instruments like Fermi;
for $\nu = 2$ the data from PKS\,2155 give $E_{\rm LIV} >  10^{-9}~E_P$.

Lorentz invariance violation would also change the transparency of the Universe.

\subsection{A win-win game: propagation and cosmology}

If no anomalous physics, an
determination of cosmological parameters
will be possible \cite{bla2005,dom2013}.

A measurement of the   Hubble constant, $H_0$ can be derived from the $\gamma$-ray attenuation observed in the spectra of  sources produced by the
interaction of extragalactic  photons with the photons of the EBL (see Equations {eq:comptau,eq:padmanabhan-diff}.

The authors of \cite{dom2013} obtain, for a $\Lambda$CDM cosmology,  $H_{0}=71.8_{-5.6}^{+4.6}({\rm stat})_{-13.8}^{+7.2}({\rm syst})$~km~s$^{-1}$~Mpc$^{-1}$.
With more data at HE (to constrain the spectral fits) and VHE (to determine well the absorption) it will be possible in the future to
 constrain other cosmological parameters, such as the total dark matter density $\Omega_{m}$ and the dark energy contribution.


\section{A SHOPPING LIST}

We list here some relevant  phenomena   in addition to the ``core fundamental science"  topics that we described in the previous sections, for which gamma-ray astrophysical detectors (IACTs in particular) could give a relevant contribution \cite{astrogam,doro}. 
\begin{itemize}
\item {\bf{Tau-Neutrino searches.}}
 Astrophysical $\tau$ neutrinos have not yet been discovered in cosmic neutrino detectors, however, they
 may be observable with IACT through a phenomenon called Earth-skimming taus \cite{7,8}. If
 a $\nu_\tau$ crosses the right amount of ground (the Earth crust, or water), of the order of few tens of
 km, tau-leptons can be generated in the process $ \nu_\tau \to \tau X$. If the tau-lepton later on emerges
 from the medium, it can create an atmospheric shower seen by a Cherenkov telescope.
 \item {\bf{Magnetic Monopoles.}}
 When a magnetic monopole crosses the Earth atmosphere, it will produce a huge flux of Cherenkov photons, thousands times than those produced by a gamma rays, which could be detectable by IACTs. A huge radiation should also be observed by EAS detectors based on water Cherenkov pools.
\item {\bf{Antiquark matter.}} Some theories
predict that during baryogenesis,  antimatter  was confined into very high dense states of
quark plasma by the formation and  collapse of domain walls in the existing quark-gluon
plasma \cite{18,19}. Such aggregations, called ``quark nuggets'', are similar to  strangelets \cite{20}) and they would be composed of a huge number of antiquarks (or quarks), O$(10^{25} - 10^{35})$. They are likely to have survived    in the intergalactic medium, and  are expected to initiate extended atmospheric
 showers; however
they would not decay in the atmosphere, and their speed would be smaller than that of cosmic
 rays, typically of the order of the Galactic velocities. Their passage would then be seen as a ``stripeÓ on the
 camera of IACT telescopes.
		\item {\bf{Primordial Black Holes.}} 
Primordial black holes (PBHs), possibly created  in the Early
 Universe, before the first stars formed, might have a wide range of possible masses  from the Planck mass to $10^5 M_\odot$  \cite{22}. As  time passes, a PBH increases its temperature and
 radiates energy, and in a final phase the BH evaporates  \cite{23}. The gamma-ray emission would
 be stable for most of the time, while an exponential increase in the last minutes to seconds to the
 evaporation is predicted.
 PBH evaporation could be therefore appear as short bursts of emission randomly in the FOV of
 an IACT regular observation.
\item {\bf{Electromagnetic counterparts of gravitational wave events and of neutrino bursts.}} The first detections of GW signals from binary black hole (BH-BH) mergers, observed by  Advanced LIGO \cite{Abbott2016a}, marked the onset of the era of GW astronomy. The next breakthrough  will be the observation of their  electromagnetic counterparts, which will characterize the progenitor and its environment. Emission up to the MeV energy range can be expected.
MeV detectors  will be crucial for the identification and the multi-wavelength characterization of the GW progenitor and of its host galaxy; they could provide the first direct evidence that binary systems are the progenitors of short GRBs.
Another important topic in the new astronomy will be neutrino astrophysics. Although astrophysical neutrinos have been detected by IceCube  \cite{astronu}, no significant cluster (in space or in time) has been found yet. MeV gamma-ray detectors could locate the emitters \cite{deaastro}.
\item {\bf{New heavy particles: top-down mechanisms.}}
Extragalactic gamma-ray emission could originate in decays of exotic particles in the early
Universe. The energy spectrum of this component should be different from the AGN contributions.
Bounds on long-lived relics have been derived using EGRET and
COMPTEL observations of the diffuse $\gamma$-ray background. 
Many models predict 
long-lived relics that may or may not be dark matter candidates. Long lifetimes for heavy relics,
larger than the age of the Universe, may arise in models with symmetry breaking at short distances.
Examples of such models are technibaryons in technicolor models or R-parity violating SUSY.
\end{itemize}

\section{FUTURE DETECTORS AND FUNDAMENTAL PHYSICS}
The future is bright for VHE gamma astrophysics 
\cite{deangelisfuture}.

In the immediate future, the Cherenkov Telescope Array CTA \cite{CTA} will provide  an order of magnitude improvement in sensitivity over existing IACTs (Figure \ref{fig:Sensitivities}).
 Different modes of operation will be possible: deep field observation; pointing mode; scanning mode. Correspondingly, different studies in fundamental physics will be possible.
CTA is expected to surpass in performance the present IACTs around 2020.

Besides new IACTs, there is a lot of activity on the development of EAS detectors, since they are relatively cheap (of the order of 30 MEUR) and we know now what is the correct size for the detection of a relevant number of sources.
\begin{itemize}
\item An upgrade of the HAWC high-energy gamma-ray observatory with a sparse array of small
outrigger tanks is in progress \cite{hawcp}.  
A gain of 3-4 in
sensitivity for gamma rays above 10 TeV can be obtained over what is presently achieved, and such a detector can be built in 2-3 years. Funds are already available.
\item
LHAASO (Large High Altitude Air Shower Observatory) is planned to  be located at about 4400 m asl and latitude $30^{\circ}\,$N in the Daochen site,  China.
It will include
 a close-packed, surface water Cherenkov detector facility with a total area of 90,000 m$^2$ (LHAASO-WCDA), four times that of HAWC.
A configuration corresponding to 25\% of the final detector might be ready by 2019 - 2020. 
\item 
HiSCORE (Hundred*i Square-km Cosmic ORigin Explorer),
located in the Tunka Valley near Lake Baikal at about 3200 m asl and latitude $51^{\circ}\,$N,  will consist of an array of wide-angle ($\Omega\sim$ 0.6-0.85 sr) light-sensitive detector stations, distributed over an area of the order of 100 km$^2$. 
The primary goal of this non-imaging Cherenkov detector is gamma-ray astronomy in the 10 TeV to several PeV range. A prototype array of 9 wide-angle optical stations, spread on a 300$\times$300 m$^2$ area, has been deployed since October 2013, and technical tests are underway. 
A 1 km$^2$ engineering array has been deployed and is in commissioning.

\item {Several proposals of EAS experiments in the Southern hemisphere}
 are being formulated now, which would be the key to study the VHE emission from the Inner Galaxy. HAWC has demonstrated that present EAS technology can be competitive with the technology of large-area coverage by small Cherenkov telescopes.
In particular among the proposals are:
\begin{itemize}
\item a Southern site for HAWC, HAWC-South, to be located in the Atacama desert \cite{hawcs};
\item LATTES, a hybrid water Cherenkov-RPC detector to be located in Southern America \cite{lattes}.
\end{itemize}

\end{itemize}

What will be the future of gamma-ray astrophysics in a longer term (say, 15 years from now), and what the consequences for fundamental physics with cosmic gamma rays? besides CTA,  we can expect:
\begin{itemize}
\item a major detector in the MeV-GeV region.  The region below $\sim$30 MeV has important implications on the science at the TeV and above, since a good knowledge of the spectra in the MeV region can constrain the fit to the emitted spectra at high energies, thus allowing:
\begin{itemize}
\item to evidence  contributions from new physics (dark matter in particular);
\item to estimate cosmological absorption (due to EBL or to possible interactions with axion-like fields \cite{darma}).
\end{itemize}
Other important aspects are:
\begin{itemize}
\item to study antimatter, in particular through detailed measurement of the 511 keV line;
\item to understand nuclear effects, evidencing the emission lines of nuclei and thus studying the chemical evolution of stars and galaxies, and the effect of collisions of outflows on molecular clouds.
\end{itemize}
On top of this, The 0.3-30 MeV energy range is important per se, but experimentally difficult to study. It requires an efficient instrument working in the Compton regime with an excellent background subtraction, and possibly with sensitivity to the measurement of polarization.
Since COMPTEL, which operated two decades ago, no space instrument obtained extra-solar gamma-ray data in the few MeV range; now
we are able to build an instrument one-two orders of magnitudes more sensitive than COMTEL based on Silicon technology, state-of-the-art analog readout, and efficient data acquisition. 

Two proposals (e-ASTROGAM \cite{astrogam} and COMPAIR \cite{compair}) have been made to ESA and NASA respectively, and convergence is likely for an experiment to be launched between 2025 and 2029.
\item A large field-of-view VHE detector in the South, which  might give substantial input with respect to the knowledge of the gamma sky, transients, and of possible PeVatrons in the Galaxy, and complement  CTA. Several proposals are being formulated now, and they should join. A large detector in Southern America could compete in sensitivity with the small-size telescopes of CTA-South already at 100 TeV, offering in addition a serendipitous approach.
\end{itemize}

\section{HIDDEN TREASURES}

Some of the best treasures
can be hidden in the least
know region of the gamma
spectrum: the MeV region (Figure \ref{fig:egb}).
\begin{itemize}
\item DM could still be light
\item Antimatter excess
manifests itself in the MeV
region
\item The MeV region in the
Inner Galaxy is presently
not well understood
\item The physics of gamma-ray
emission needs to know
well the transition from
atomic to nuclear
processes, and from
nuclear to QCD/HEQED
\item Understanding
nucleosynthesis needs
measuring MeV lines.
\end{itemize}

\section{SUMMARY}

There is a clear interplay between gamma-ray astrophysics and fundamental physics;
this model of cooperation has worked well, and will certainly work well in the future. Main subjects are:
\begin{itemize}
\item  The study of the propagation of photons over cosmological distances.
\item Search for dark matter and new particles in photon spectra.
\item Study of physics in extreme environments (cosmic rays, jet properties, É; only
marginally covered in this presentation).
\end{itemize}

High-energy gamma-ray  astrophysics is exploring regions beyond the reach of accelerators.
A ``simple'' extension of present Cherenkov detectors is in progress: CTA, and promises new discoveries in the next years, but also large field-of-view ground-base detectors are going to be built in the next years
 (the upgraded HAWC, LHAASO, new detectors in the Southern hemisphere). 
New physics which might also come from satellites sensitive to photons in the MeV range (e-ASTROGAM, COMPAIR).


\section{ACKNOWLEDGMENTS}
I tkank Felix Aharonian, Frank Rieger and Werner Hofmann for inviting me to this very stimulating conference and for many discussions. I have partly followed Carr et al. \cite{carr} on the 
description of photon signatures in DM annihilation, and Doro \cite{doro} on the search for rare events.


\nocite{*}
\bibliographystyle{aipnum-cp}%

\end{document}